# Emerging Ferroelectric Domains: Stacking and Rotational Landscape of MoS$_2$ Moiré Bilayers

Anikeya Aditya[1], Ayu Irie[2], Nabankur Dasgupta[1], Rajiv K. Kalia[1], Aiichiro Nakano[1], and Priya Vashishta[1*]

[1] *Collaboratory for Advanced Computing and Simulation, Department of Chemical Engineering & Materials science, Department of Computer Science, Department of Physics & Astronomy, and Department of Quantitative & Computational Biology, University of Southern California, Los Angeles, CA 90089-0242, USA*
[2] *Department of Physics, Kumamoto University, Kumamoto 860-8555, Japan*

*Corresponding author: priyav@usc.edu

## Abstract

The structures and properties of moiré patterns in twisted bilayers of two-dimensional (2D) materials are known to depend sensitively on twist angle, yet their dependence on stacking order remains comparatively underexplored. In this study, we use molecular dynamics simulations to systematically investigate the combined effects of stacking order and rotation in MoS$_2$ bilayers. Beginning from five well-established high-symmetry bilayer stackings, we apply twist angles between 1° and 120° to the top layer, revealing a variety of relaxed moiré structures. Our results show that the initial stacking significantly influences the moiré domain configurations that emerge at a given twist angle. While all five stacking orders are metastable without twist, they form two moiré-equivalent classes—AA/AB and AA′/A′B/AB′, *i.e.*, for a given twist angle, structures within each class relax to the same moiré configuration. Specifically, initial AA and AB stackings give rise to triangular ferroelectric domains near 0±3°, while AA′, A′B, and AB′ stackings produce triangular ferroelectric domains near 60±3°. At precisely 60° and 120° twists, the bilayers relax to into pure high-symmetry stackings, highlighting the rotational relationships between these configurations and explaining the shift of 60° in the ferroelectric rotational range. These findings demonstrate the critical role of stacking order in governing the rich moiré landscapes accessible in twistronic systems.

# Introduction

Bilayer 2D materials have captivated researchers for their ability to form moiré superlattices, driving advancements in twistronics.[1-9] The interference patterns arise when two lattices overlap with structural differences or angular misalignment, producing long-wavelength superlattices with unique electronic and optical characteristics. Moiré structures exhibit extraordinary properties, including superconductivity,[10, 11] excitons with distinct optical signatures,[13-17] correlation-driven Mott insulating states,[18, 19] and quantum anomalous Hall effects.[20, 21] Extensive studies on these exotic phenomena have been conducted across various 2D material systems, including graphene,[22, 23] hexagonal boron nitride (h-BN),[24] and transition metal dichalcogenides (TMDCs),[20, 25-33] demonstrating distinct electrical and optical modulations by moiré superlattices. For instance, twisted $MoS_2/WSe_2$ heterobilayers exhibit moiré patterns from lattice mismatch, with periodic atomic registry variations.[30] The electronic structure at critical points is influenced by interlayer coupling, which maintains the characteristics of a direct gap semiconductor even though the valence and conduction band edges are situated in separate layers.[30] Another notable example is the emergence of ferroelectric domains in twisted $WSe_2$ homobilayers, whose polarization is imprinted on the plasmonic response of adjacent graphene monolayers.[26] The work demonstrated how ferroelectric domains in the moiré structure can effectively modulate optoelectronic properties of proximate graphene. The exotic properties of moiré superlattices hold promise for applications in quantum information,[34, 35] optoelectronics,[36, 37] and strain engineering,[38] enabling advancements in quantum states, light-matter interactions, and tailored material functionalities.

Lattice reconstruction plays an essential role for the structures and properties of bilayer 2D materials, where atomic positions shift to minimize energy, thereby forming discrete commensurate domains delineated by narrow domain walls or solitons, rather than a smooth registry.[25, 28, 39-41] For instance, twisted heterostructures of $MoSe_2/WSe_2$ and $MoS_2/WS_2$ undergo considerable atomic reconstructions at twist angles ≤1°, resulting in a superlattice of discrete domains separated by domain walls rather than continuously varying moiré patterns.[42] Another important issue in moiré bilayers is stacking order. Experimental and theoretical analyses show triangular domains for AB stacking and hexagonal domains for AA' stacking, along with modulations in electronic band edges and conductivity domains.[29] Simulations of twisted bilayer $MoS_2$ reveal distinct structural regimes driven by twist angles.[29] At 13°–47°, the layers act as rigidly twisted monolayers, while angles <3° or near 60° form domain-soliton networks with triangular stacking domains and strain solitons. These reconstructions flatten electronic features, such as Dirac cones and Kagome bands, underscoring the critical role of stacking sequence.

TMDC homobilayers exhibit five high-symmetry stacking orders as shown in Fig. 1, among which AB and AA' stackings (also referred to as 3R and 2H stackings, respectively) are particularly noteworthy for their low energies. Among them, AB stacking stands out for its ferroelectric nature due to out-of-plane polarization.[43, 44] The role of stacking order in moiré superlattices remains underexplored despite its critical influence on symmetry and energetics. Among few pioneering studies, experiments by Weston *et al*. on twisted bilayers of $MoS_2$ and $WS_2$ revealed that stacking order profoundly shapes domain morphology.[45] Their scanning transmission electron microscope (STEM) images showed triangular domains in bilayers twisted 0.65° from 3R stacking and hexagonal domains in bilayers twisted 0.25° from 2H stacking. These findings underscore the need to introduce stacking order as a design parameter in twistronics.

In this work, we employ molecular dynamics (MD) simulations to investigate the formation of moiré patterns in relaxed twisted bilayer $MoS_2$, starting from various initial stacking configurations. We systematically rotate bilayers starting with initial five stacking, AA, AB, AA′, A′B, and AB′, by angles between 1° and 120° and analyze the stacking domain populations of the relaxed structures. Our results show that the initial stacking influences the nature of the moiré domains formed at a given twist angle. All five starting stacking configurations can give rise to both ferroelectric triangular domains and non-ferroelectric hexagonal domains, depending on the twist angle. Initial 3R-like stackings (AA and AB) produce triangular ferroelectric domains near 0 ± 3°, while 2H-like stackings (AA′, A′B, AB′) generate similar ferroelectric domains near 60 ± 3°. This shows that moiré structures starting within 3R and 2H families are moiré equivalent. Furthermore, at a 60° rotation, the system transforms from a 3R stacking to a 2H stacking structure and vice versa, revealing the rotational relationship between the two classes and the 60° shift in the range of triangular domains.

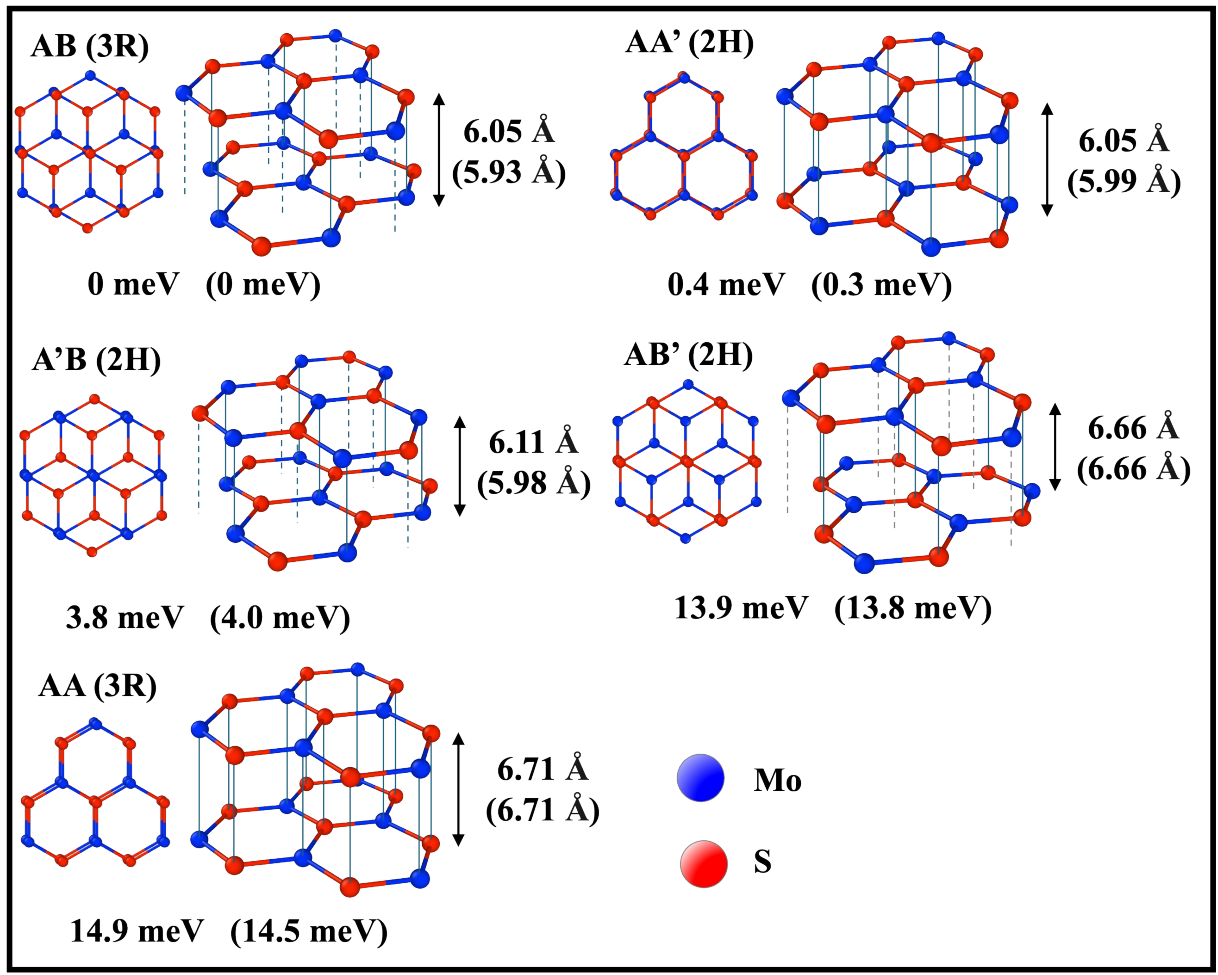

**Fig. 1.** Top (left) and perspective (right) views of atomic structures for five different stacking orders of bilayer $MoS_2$. The calculated interlayer distance (defined between Mo atoms on adjacent layers, in Å) and the relative energy compared to the most stable AB stacking (in meV) are shown. Molecular-dynamics results from this study are presented without parentheses, while first-principles values from Ref. 12 are given in parentheses. Blue and red colors indicate molybdenum and sulfur atoms, respectively.

## Results

Bilayer MoS$_2$ without any twist has six known stacking formations: AA, AB, BA, AA', A'B, AB', as shown in Fig. 1. In fact, five stacking because the AB and BA stackings have the same atomic arrangement but flipped about *z* direction, hence they are equivalent in the absence of external field. Liu *et al*. investigated five stacking – AA, AB, AA', A'B, AB' using density functional theory (DFT) and reported the interlayer distances (ILD) and relative energies (ΔE) using both LDA and PBE-D functionals.[12] We calculated the ILD and ΔE using MD, and the values are in good agreement with the LDA values shown in Fig. 1. The stackings are arranged in ascending order of relative energies. The stackings are often grouped into two groups: parallel containing AA and AB (BA) and antiparallel containing AA', A'B, and AB'.[46] The parallel stacking is also referred to as 3R and antiparallel stacking as 2H. The relative stability of different stacking at room temperature was investigated by heating the system to ~300K in the microcanonical (NVE) ensemble. We found that AB/BA, AA' and A'B remain stable at room temperatures, while AA and AB' transform to AB and A'B at room temperatures. More detail on the finite temperature stability and energetics are given in Supplementally Information (Fig. S1). The optimal path to move from one type of stacking to another within the same group is investigated in Ref. 4.

### Effects of initial stacking order

In literature, twisted bilayer MoS$_2$ research is focused on the formation of triangular 3R domains or hexagonal 2H domains. To investigate the impact of the initial bilayer configuration on the formation of moiré domains, we took the five bilayers and rotated them by 1° and 59°. Figure 2a shows the characteristics of the stacking domains that emerge in bilayer MoS$_2$ when identical twist angles are applied, beginning from different initial stacking configurations. Initial 3R stacking (AA and AB) bilayer with 1° twists form triangular AB and BA domains separated by dislocation lines and high energy AA stacking at the triangle vertices. On the other hand, initial 2H stacking forms hexagonal-like domains with low energy AA' stacking filling the domains and A'B present in the regions between three AA' domains with high energy AB' present at the vertices; unknown X stacking is present at the boundaries between these vertices. Figure 2a shows the effect of twist by 59° on the five bilayers; the initial 3R stacking now forms hexagonal domains, and the 2H domains form triangular domains. Figure 2b and 2c show the zoomed in view of the presence of high energy AA and AB' at the vertices.

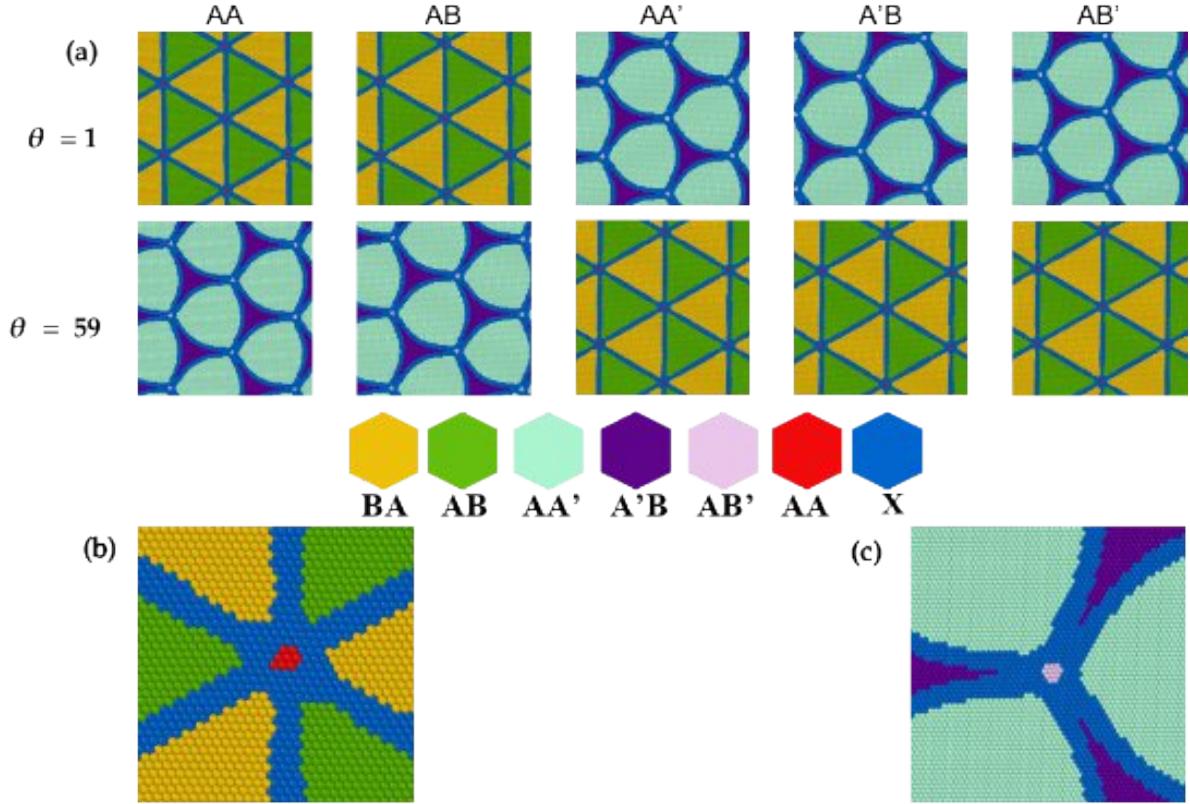

**Fig. 2.** Stacking domains formed when the same twist is introduced in the bilayer (BL) MoS$_2$, starting from different initial stacking orders. When the twist is closer to 0°-1°angle, the BLs initially in AA and AB stack form triangular domains of AB and BA stacking separated by dislocation lines and high energy AA stacking at the triangle's vertices. The BLs initially in AA', AB', and A'B form hexagonal domains of AA' stacking with AB' stacking at three vertices and regions of A'B on the other three vertices. The blue color atoms correspond to unknown (X) stacking. The situation reverses when the rotation is closer to 60°-59° angles; the AA and AB bilayers form hexagonal domains, while AA', AB', and A'B form triangular domains. Panels b and c show that the high-energy AA (red) and AB'(pink) domains exist at the vertex of these domains surrounded by X (blue) atoms.

To further investigate the populations of different stacking types in the twisted bilayers, the five highly symmetric initial configurations described in Fig. 1 were rotated incrementally from 1° to 120° in steps of 1°. After structural relaxation, we analyzed the stacking populations per atom of each configuration. Figure 3 shows the percentage of atoms in 3R (*i.e.*, AA or AB/BA) stacking as a function of twist for the five different initial stacking bilayers. For the bilayers initially in BA and AA stacking, the population of 3R stacking decreases rapidly between 0° and 10°, then it plateaus in the range (15°, 22°), and then decreases to zero by 40°. Then, it shows the mirror behavior increasing from 80° onwards. At 30° and 90°, the population of 3R stacking is equal for all the bilayers.

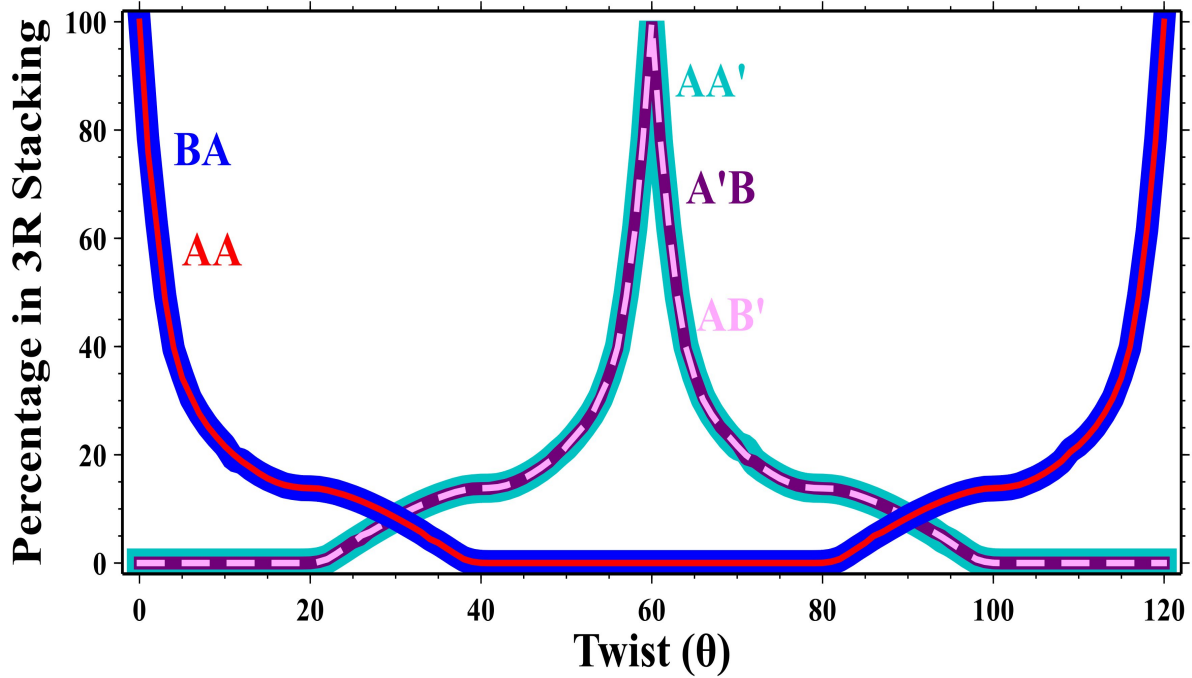

**Fig. 3.** Percentage of atoms exhibiting 3R stacking sequences (AB, BA, and AA) for various initial stacking orders. The solid blue and red lines represent systems initially in BA and AA stackings, respectively, prior to applying the twist. The cyan solid, purple solid, and pink dotted lines correspond to systems initially configured in AA', A'B, and AB' stackings, respectively.

Figure 4 shows the percentage of atoms in 2H stacking as a function of twist for the five different initial stacking orders. For bilayers initially in AA', A'B, or AB' stacking, the population decreases rapidly between 0° and 10°, then it plateaus in the range (15°, 22°), and then decreases to zero by 40°. Then, it shows the mirror behavior increasing from 80° onwards. At 30° and 90°, the population of 2H stacking is equal for all the bilayers. This behavior mirrors what we see in Fig. 3 for the 3R stacking. Moreover, at 30° and 90°, the population distribution in 3R and 2H stacking is the same. The 30° corresponds to the quasicrystal regime where no periodic unit cell can exist.

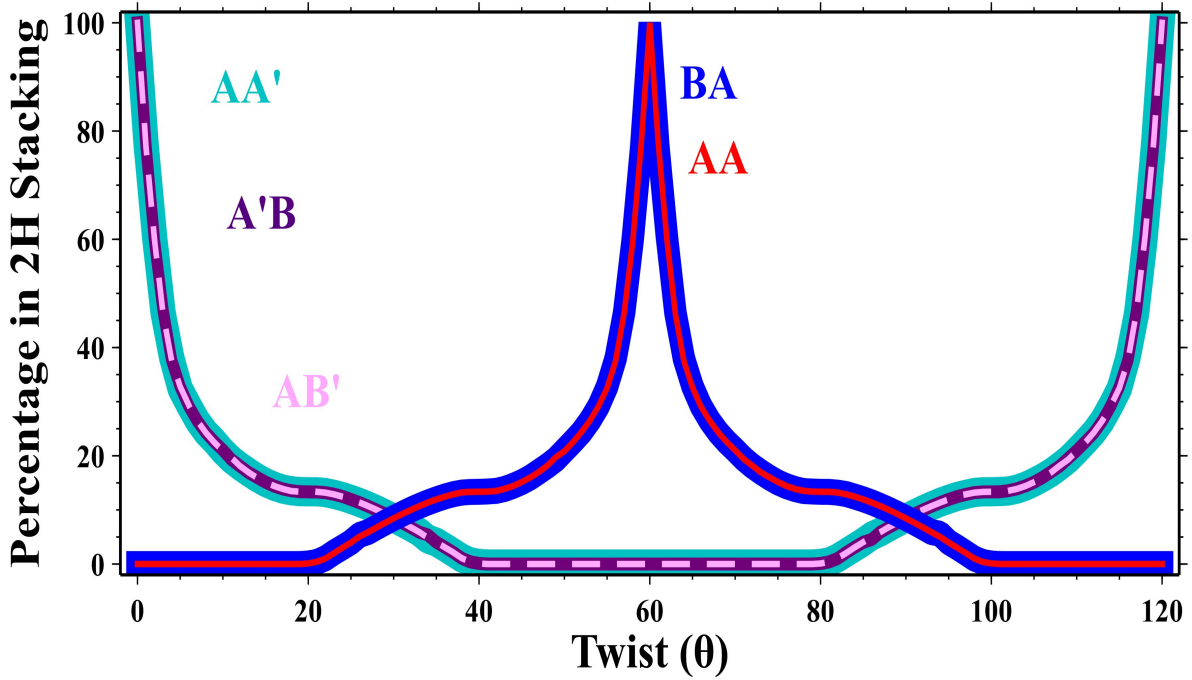

**Fig. 4.** Percentage of atoms exhibiting 2H stacking sequences (A'B, AB', or AA') for various initial stacking configurations. The solid blue and red lines represent systems initially in BA and AA stackings, respectively, before the twist was applied. The cyan solid, purple solid, and pink dotted lines correspond to systems initially configured in AA', A'B, and AB' stackings, respectively.

Moiré $MoS_2$ bilayers form ferroelectric domains, which are characterized by triangular domains of alternating AB and BA domains separated by domain boundaries and the presence of high-energy AA stacking at the vertices of the triangular domains. The size of the AB and BA domains in ferroelectric structures is the same. To find the twist angles where these ferroelectric domains exist, the population percentage of AB and BA domains is tracked for two systems, initially in BA and AA' stacking. Only these two systems were chosen as they both are low-energy stackings, and Figs. 3 and 4 show that under twist, all the 3R and 2H systems are moiré equivalent. Figure 5 shows the population percentage of atoms in AB and BA stackings. The dashed lines represent the population percentage for AB stacking, where the green dashed line represents the system that was in BA initially, and the blue dashed line represents the system that was in AA'. The solid lines with solid circular markers represent the population for BA stacking, and the color scheme is the same for the initial BA and AA' stacking as described above. Starting with the bilayers in BA stacking (solid green line and circular marker), at 0° twist, all the atoms were categorized as being in BA stacking, at 1° twist, the percentage of BA falls to about 40%, and AB rises to 40%, and we have ferroelectric domains. This behavior continues until 3°, when the population of AB and BA domains is the same. Beyond this, the population share of AB falls off, and the domains formed are no longer clean triangular domains with alternating polarity. Similar behavior is seen between the twist range of 117° and 120°. Therefore, the range of twists where ferroelectric domains form when the initial system is in AB (3R) configuration is 0° to 3° and 117° to 120°. At 120° twist, the system transforms completely into AB stacking, which shows the rotational relationship between AB and BA stacking. As the behavior is symmetric about 60°, we can conclude that the ferroelectric range for 3R system is 0±3°. Similarly, initial twists do not produce ferroelectric domains starting with the AA' system. As shown in Fig. 5, at 57° twist, the population of AB and BA becomes the same, and triangular ferroelectric domains form between 57° and 63°, except at 60°. At 60°, the initially AA' system transforms into the AB system, revealing the rotational relationship between the two configurations.

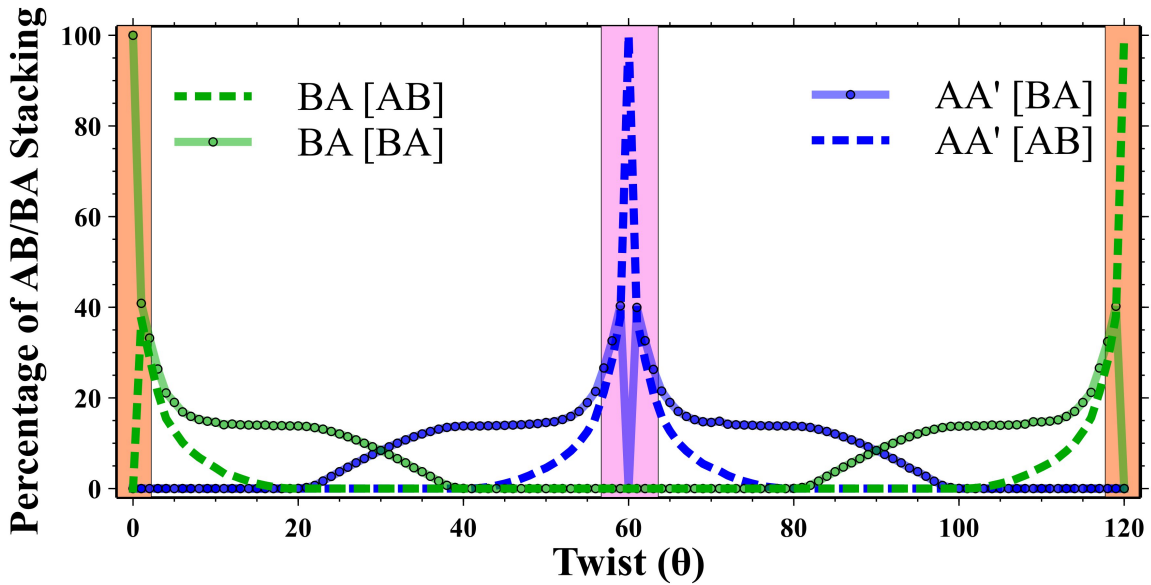

**Fig. 5.** Percentage of BA/AB stacking, which indicates the presence of ferroelectric moiré domains, as a function of twist angle. The dashed green and blue lines represent the percentage of atoms in AB stacking for systems initially configured in BA and AA' stacking, respectively. The thicker green and blue lines with circular markers represent the percentage of atoms in BA stacking for the corresponding initial stackings (BA and AA', respectively). Ferroelectric regions, resulting from twists of 0–3° and 117–120° for the initially AB (3R) stacked systems, are highlighted in orange. Similarly, ferroelectric regions emerging from twists between 57° and 63° for initially AA' (2H) stacked systems are highlighted in pink

The rotational relationships discussed above were further investigated for each initial stacking configuration, and the resulting high-symmetry structures observed at rotations of 60° and 120° are summarized in Table 1. A bilayer initially in AA stacking transforms into AA' at 60° and subsequently into BA at 120°, without reverting to AA. The AA configuration has significantly higher energy compared to AB or BA; thus, with a slight perturbation, the system preferentially relaxes into BA stacking rather than AA. Similarly, an initial BA stacking evolves into A'B and AB configurations at 60° and 120° rotations, respectively. Initial AA' stacking converts to AB at a 60° rotation and subsequently to A'B at 120°. The initial A'B stacking transitions to BA at 60° and returns to A'B at 120°. This reveals an interesting pathway from AA' → AB → A'B → BA → AA' on 60° twists. Lastly, AB' transforms into AB at 60° and AA' at 120°. Detailed analyses of these stacking transformations, including the formation of intermediate high-energy and previously unidentified stackings, are presented in supplementary Figs. S2-S6.

**Table 1.** Rotational relationship between high-symmetric stackings. Starting with different high symmetry stacking rotations by 60° and 120° produces one of the other high symmetry stacking structures. 3R class structures are italicized, and 2H structures are bold.

| **Starting ($\theta = 0°$)** | $\theta = 60°$ | $\theta = 120°$ |
|---|---|---|
| *AA* | **AA'** | *BA* |
| *BA* | **A'B** | *AB* |
| **AA'** | *AB* | **A'B** |
| **A'B** | *BA* | **AA'** |
| **AB'** | *AB* | **AA'** |

**Ferroelectric structures**

We identified angular regions spanning approximately ±3° that correspond to the formation of ferroelectric triangular domains. For bilayers initially in the 3R configuration, ferroelectric domains emerge within twist-angle ranges of 0±3°. Similarly, for bilayers initially in the 2H configuration, ferroelectric domains appear between 60±3°, except precisely at 60°, where no distinct moiré structure forms. A representative structures from these ferroelectric regions are shown in Fig. 6, accompanied by calculated dipole moments in all three spatial dimensions, aligning closely with the observed domain shapes. Figure 6a shows a bilayer initially in BA stacking with a 1° twist, resulting in triangular domains alternating between AB and BA stackings. In-plane dipoles (X and Y directions) vanish within these triangular domains but exhibit values of approximately ±5 D at domain boundaries. Conversely, the out-of-plane polarization (Z-direction) is substantial, with magnitudes of approximately ±60 D. For comparison, Figure 6b shows an initially AA' stacked bilayer twisted by 1°, forming hexagonal domains primarily consisting of AA' stacking. Dipole moments are absent within these hexagonal domains but are present along their boundaries, as indicated in the figure.

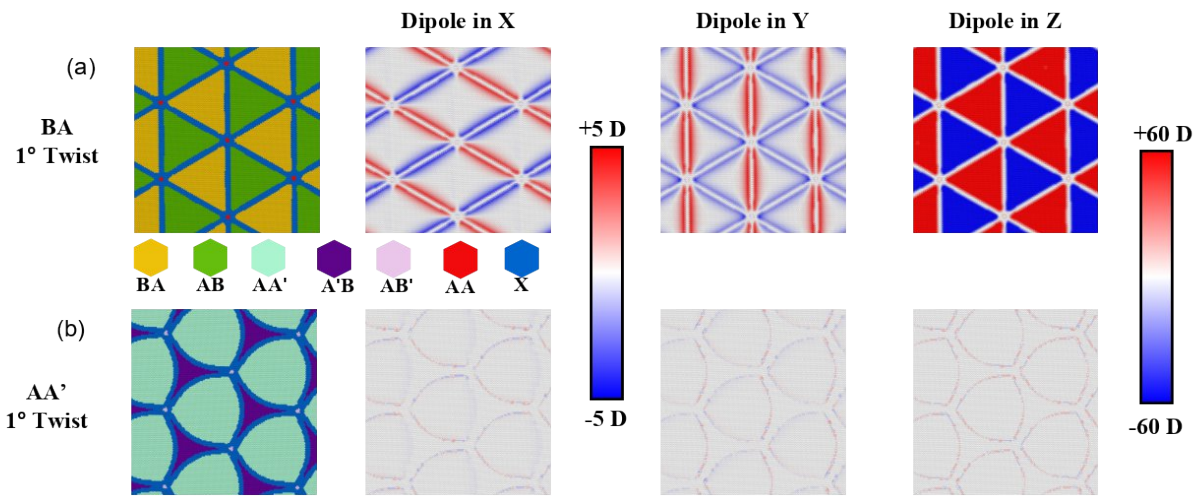

**Fig. 6.** Calculated dipole moment distribution for a moiré MoS$_2$ structure with an initial BA stacking configuration and 1° twist. The dipole moments in the *x* and *y* directions emerge predominantly at grain boundaries separating the triangular domains, while remaining negligible within the domains. In contrast, significant out-of-plane (*z*-direction) dipole moments exist within the triangular domains, with positive dipoles corresponding to AB domains and negative dipoles to BA domains. For comparison, an initially AA' stacked system with 1° twist forms hexagonal domains that exhibit no polarization.

## Conclusion

In conclusion, our analysis reveals the formation of distinct stacking domains in twisted MoS$_2$ bilayers, strongly dependent on both initial stacking orders and applied twist angles. We identified specific angular ranges where ferroelectric moiré bilayers emerge: 0±3° for systems initially in the 3R (AB) stacking, and 60±3° for those in the 2H (AA') stacking. Detailed investigation into stacking populations demonstrated clear evolution pathways among high-symmetry stacking configurations, influenced significantly by their relative energetic stability. We found that 60° rotation transforms structure from 3R to 2H stacking, explaining the shift of 60° in range of ferroelectric structures. Dipole moment analysis further confirmed polarization — markedly strong out-of-plane dipoles (±60 D) within triangular domains formed at a 1° twist from BA stacking, accompanied by in-plane polarization at domain boundaries (±5 D). In contrast, hexagonal domains dominated by AA' stacking displayed negligible internal polarization, highlighting distinct symmetry-dependent behavior. These findings underscore the critical role of initial configurations and precise twist control in engineering polarization states and ferroelectric properties in moiré superlattices.

## Methods

### Simulation schedule and force field

The simulations were performed using LAMMPS software[48] with Stillinger-Weber (SW)[46] and Kolmogorov-Crespi (KC)[49] force fields. The SW force field is used to model the interaction between atoms (Mo and S) within the same MoS$_2$ monolayer, while the KC force field is used to model the interlayer interaction. To examine the domain formation and completely control the twist angle between the monolayers, we created a circular bilayer flake with a diameter of 4,000

Å and rotated the top circular flake about the center with the desired twist angle. To prevent any slip or rotation during energy minimization, the boundary atoms within 50 Å from the circumference are not allowed to move, as shown by the red region in Fig. 7. The structure is relaxed using a conjugate-gradient energy minimization method with energy and force tolerance of $10^{-6}$ eV and eV/Å, respectively.

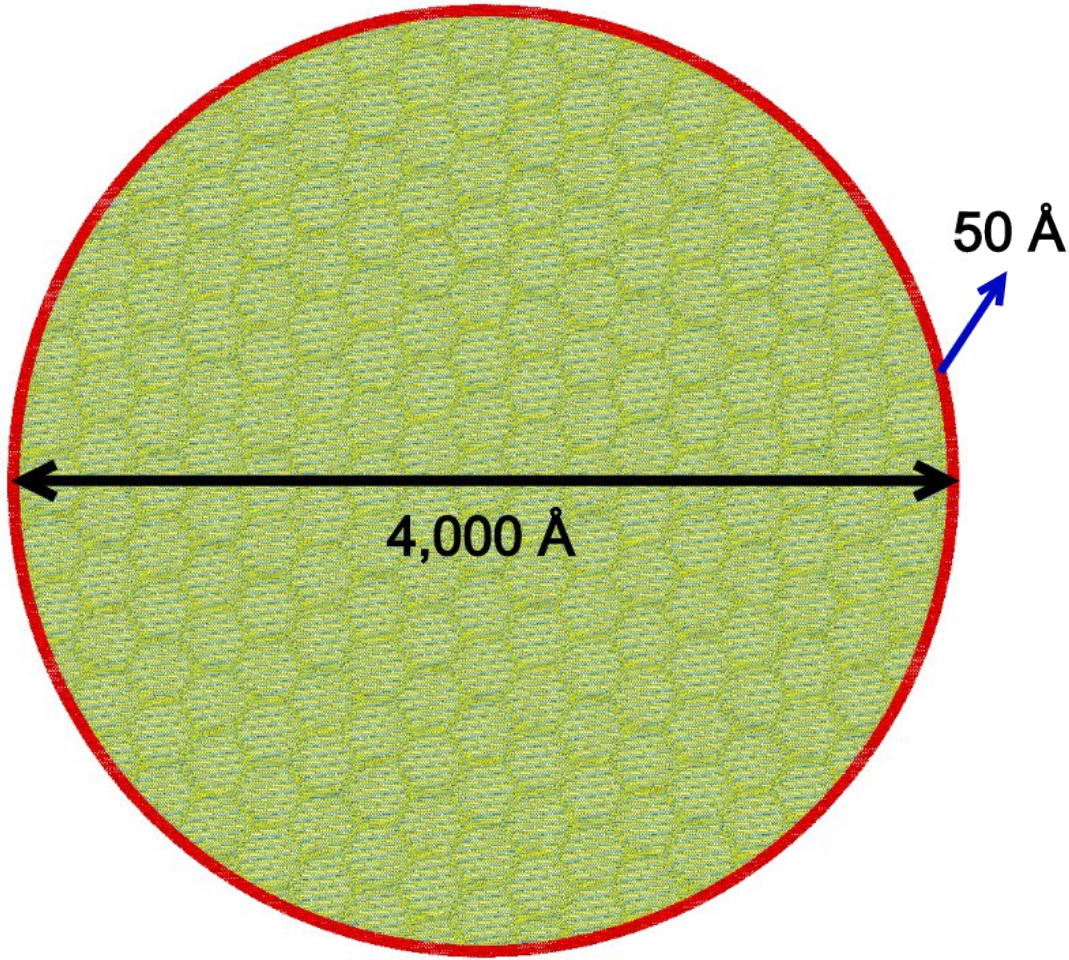

**Fig. 7.** Circular bilayer flake structure used in the simulations. The flake diameter is 4,000 Å, with a 50 Å boundary region where atoms were frozen to constrain the rotational orientation. Structural relaxation was subsequently performed using conjugate-gradient energy minimization employing Stillinger-Weber and Kolmogorov-Crespi force fields.

**Domain stacking**

Python code was developed to classify the stacking domains present in a moiré structure. The code iterates over each Mo atom in the top layer, and then identifies the three closest sulfur neighbors in the top layer. Then, for each of the selected four atoms (Mo + 3 S), a cylindrical region is created, and an index (or signature) is assigned to each of the four atoms based on the number and type of atoms below it in the cylindrical region. Each stacking type has a unique signature, and then the stacking type is assigned based on it; see Fig. 8.

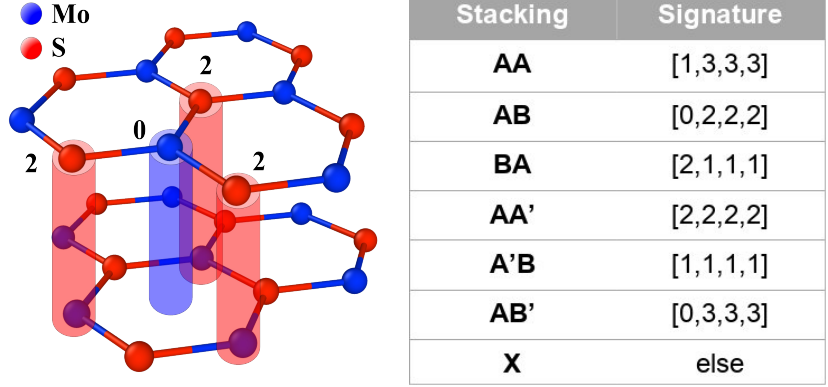

**Fig. 8.** The schematic used to assign a stacking type for each Mo atom in a given bilayer. The four selected atoms from the top layer are used to form a cylindrical region, and the number and type of atom below them inside the cylinderical region is examined. For this specific case of BA stacking, Mo has 2 sulphur atoms below it, and the three top sulphur atoms will have just one sulphur from the same monolayer below them. Hence this quartet is assigned a code of [2,1,1,1]. Similar fingerprint for other stacking is given in the table on the left. Any atom which has values diffffernet than those in the table is assgined X stacking.

**Dipole Moment Calculation**

The local dipole moment was calculated by assigning nominal charge values to Mo and S atoms based on the stoichiometric ration, using a fitting parameter $a = 1$. For each Mo atom, we defined a surrounding cylindrical region and computed the dipole moment as $\vec{D} = \sum_i q_i \vec{r}_i$, where $q_i$ and $\vec{r}_i$ are the charge and position of the i-th atom within the region. This local approach captures the dipolar character of vertically stacked configurations and allows analysis of ferroelectric domains consistent with prior studies{Vizner Stern, 2021, Interfacial ferroelectricity by van der Waals sliding} {Wang, 2022, Interfacial ferroelectricity in rhombohedral-stacked bilayer transition metal dichalcogenides} see Fig. 9. After calculating the dipole moments for each Mo atom in both layers, we averaged the values across the two layers to obtain a single representative dipole field.

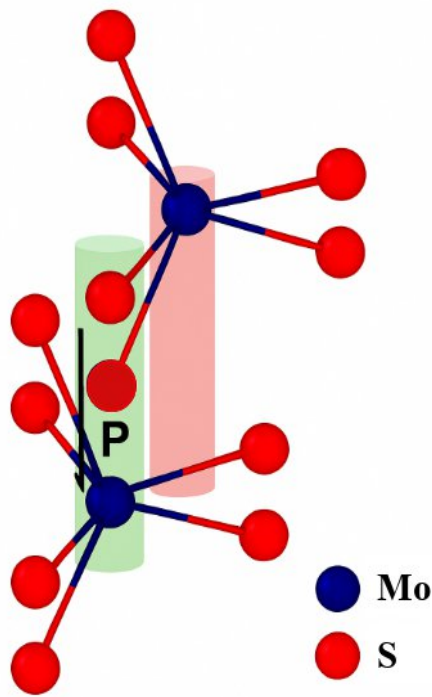

**Fig. 9.** Schematic of the local dipole moment calculation for a Mo atom in twisted bilayer MoS$_2$. Each Mo atom is surrounded by a cylindrical region (shaded in green and red for bottom and top layers, respectively), within which the dipole moment is computed as $\vec{D} = \sum_i q_i \vec{r_i}$, where $q_i$ and $\vec{r_i}$ are the charge and position of atoms (S in yellow, Mo in cyan) relative to the central Mo atom. The resulting dipole vector $\vec{P}$ points along the vertical stacking direction. Dipole values are computed for each Mo atom and averaged across the two layers to characterize local ferroelectric domains.

## Supporting Information:

Stability and energetics of high-symmetry stackings at room temperature (Figure S1), stacking population distribution starting from different stacking orders- AA (Figure S2), BA (Figure S3), AA' (Figure S4), AB' (Figure S5), A'B (Figure S6) (PDF)

## Acknowledgment

This research is supported by National Science Foundation, Future of Semiconductors Program Award Number (FAIN): 2235462.

Supplementary Information

for

# Emerging Ferroelectric Domains: Stacking and Rotational Landscape of MoS$_2$ Moiré Bilayers

Anikeya Aditya[1], Ayu Irie[2], Nabankur Dasgupta[1], Rajiv K. Kalia[1], Aiichiro Nakano[1], and Priya Vashishta[1*]

[1] *Collaboratory for Advanced Computing and Simulation, Department of Chemical Engineering & Materials science, Department of Computer Science, Department of Physics & Astronomy, and Department of Quantitative & Computational Biology, University of Southern California, Los Angeles, CA 90089-0242, USA*

[2] *Department of Physics, Kumamoto University, Kumamoto 860-8555, Japan*

*Corresponding author: Priya Vashishta, email: priyav@usc.edu

**Stability and Energetics of High-Symmetry Stackings at Room Temperature**

All five stacking types were initially relaxed using the conjugate gradient (CG) method, followed by heating to the temperature of 300 K using molecular dynamics (MD). After initial temperature scaling, the systems were equilibrated for 100,000 MD steps in the microcanonical (NVE) ensemble with the unit time step of 0.25 fs, and the potential energy per atom was averaged over the final 2,000 steps. During this process, AA stacking transformed to AB, BA remained stable as BA, AA′ remained as AA′, and both A′B and AB′ relaxed to A′B. The BA configuration exhibited the lowest potential energy and was used as the reference (zero-energy) state for calculating the relative energies of the other equilibrated configurations. In the figure, the relative energies (in meV per atom) are indicated in parentheses.

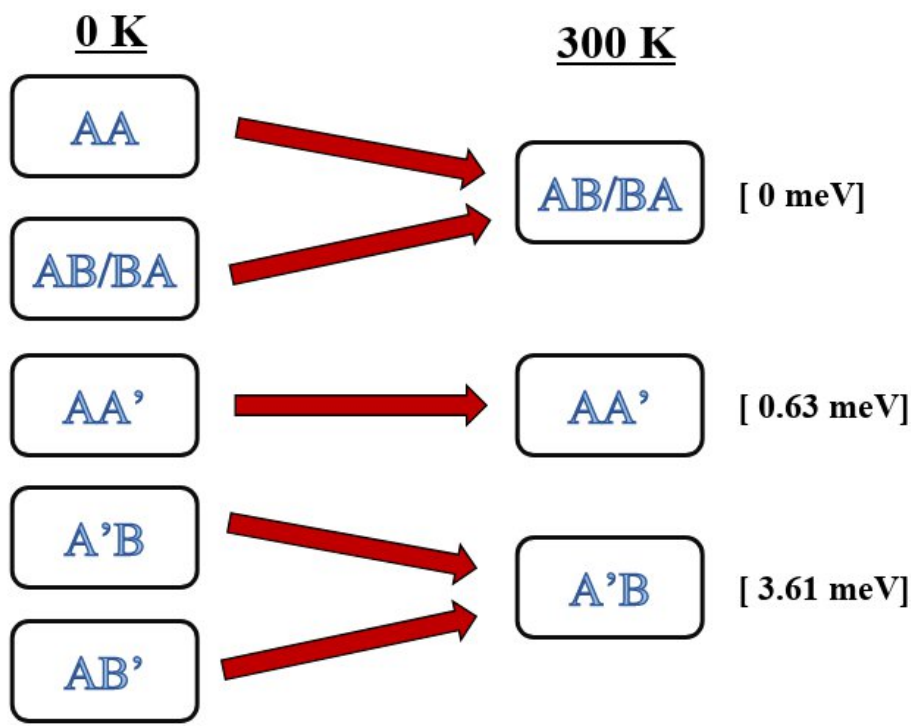

**Fig. S1.** Stability and energetics of high-symmetry stacking configurations at room temperature. The relative energies (in meV per atom) are indicated in parentheses.

**Stacking Population Distribution Starting from Different Stacking Orders**

Figures S2 to S6 show how the percentage of atoms in different stacking arrangements changes with the twist angle between the layers. The red line represents the percentage of atoms in AA stacking. The green lines show the percentage of atoms in AB stacking. The ~~yellow~~ brown line represents BA stacking, the teal line indicates AA' stacking, and the black line shows AB' stacking. The purple line represents A'B stacking. The symbol X refers to atoms that do not fit into any known stacking arrangement or have an unknown stacking type. The figures reveal rotational relationship between the high symmetry stackings at exactly 60° and 120° rotations, and a pathway for such transformations.

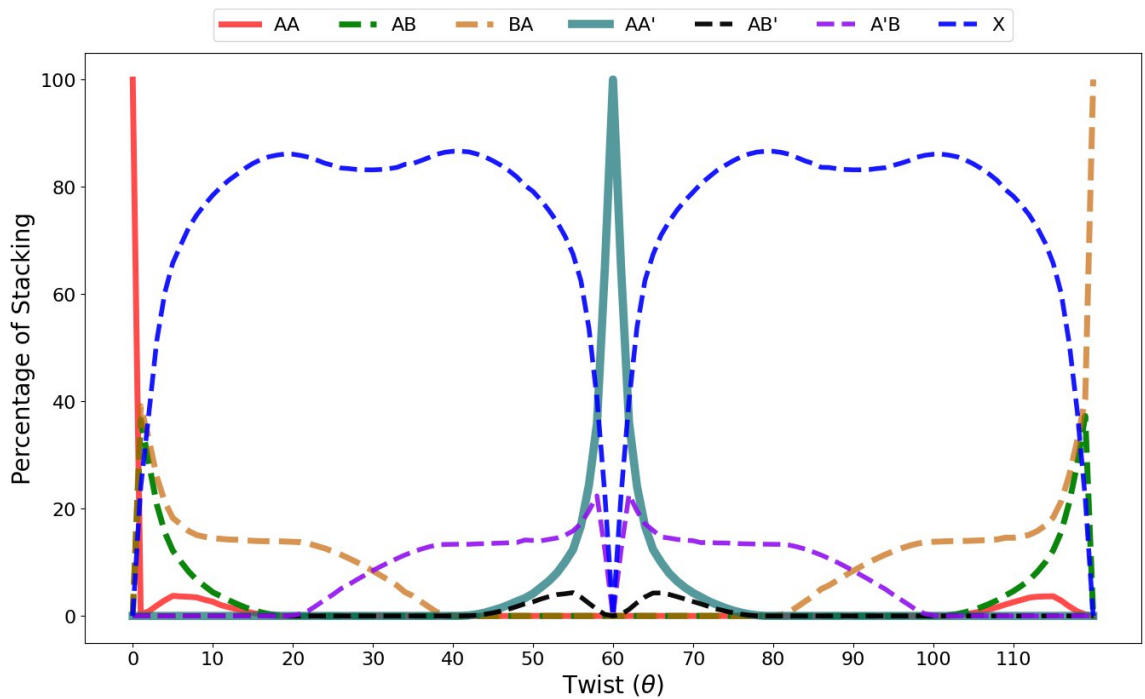

**Fig S2.** Starting with AA stacked MoS$_2$ bilayers, twists are introduced between the bilayers and the percentage of atoms in different stacking types are calculated. At 60° rotation, the bilayers transform to AA' and BA stacking at 120°.

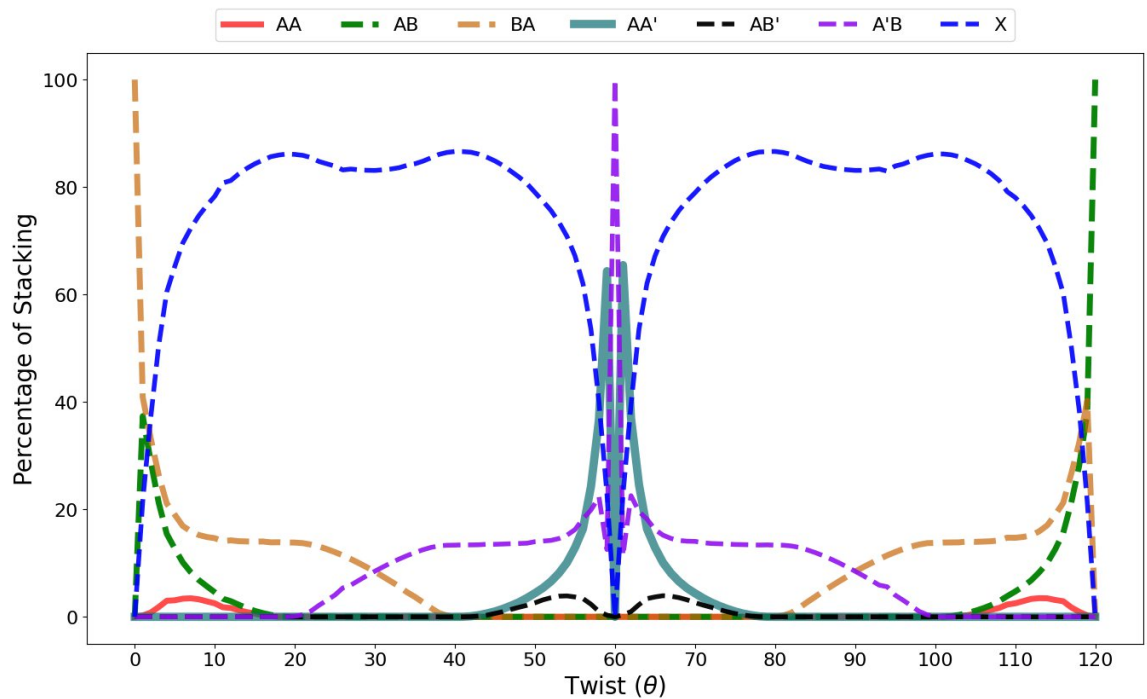

**Fig. S3.** Starting with BA stacked MoS$_2$ bilayers, twists are introduced between the bilayers. At 60° rotation, the bilayers transform into A'B and at 120° rotation it transforms into AB.

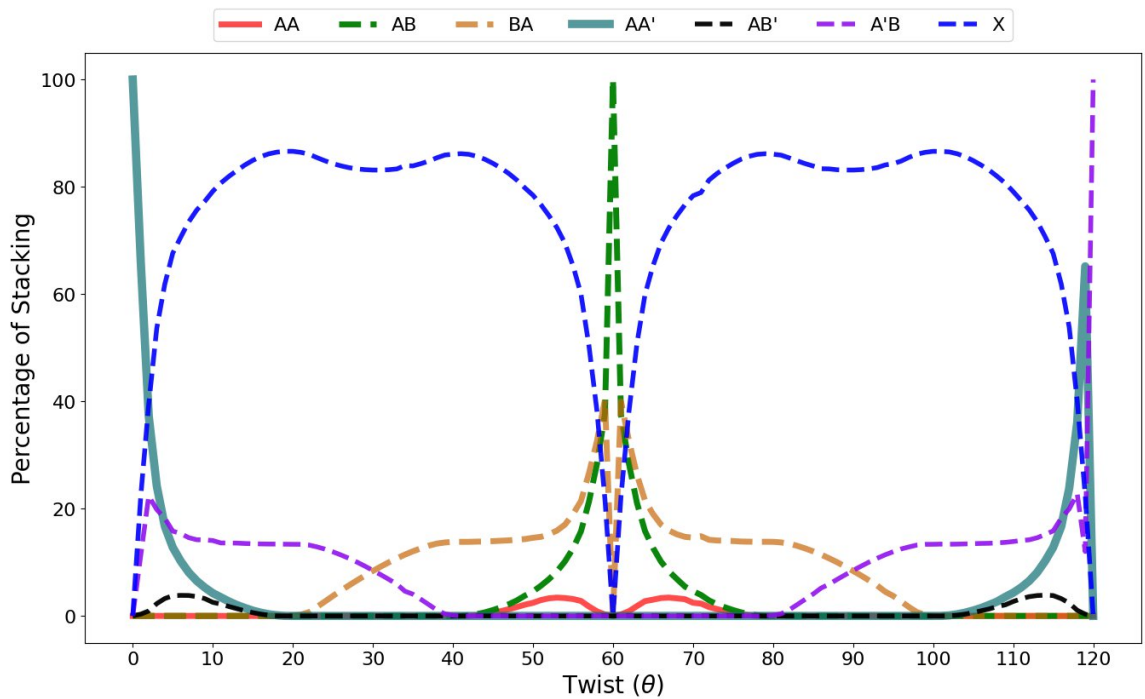

**Fig. S4.** Starting with AA' stacked MoS$_2$ bilayers, twists are introduced between the bilayers. At 60° rotation, the bilayers transform to AB and at 120° into A'B.

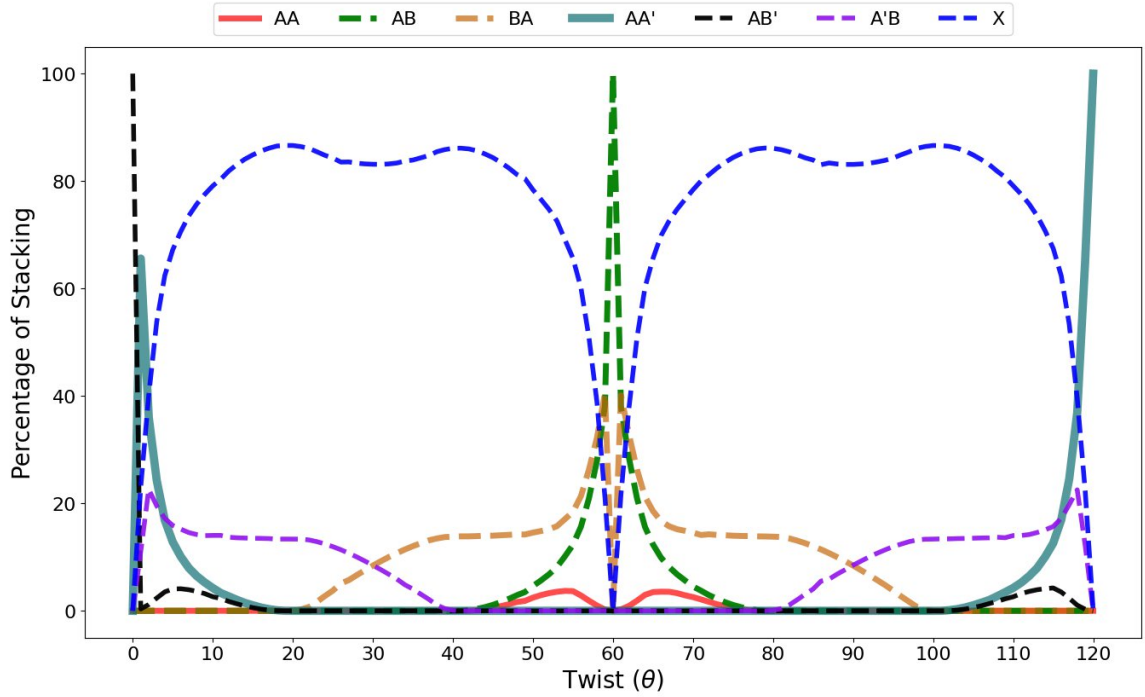

**Fig. S5.** Starting with AB' stacked MoS$_2$ bilayers, twists are introduced between the bilayers and the percentage of atoms in different stacking types are calculated. At 60° rotation, the bilayers transform into AB and into AA' stacking at 120°.

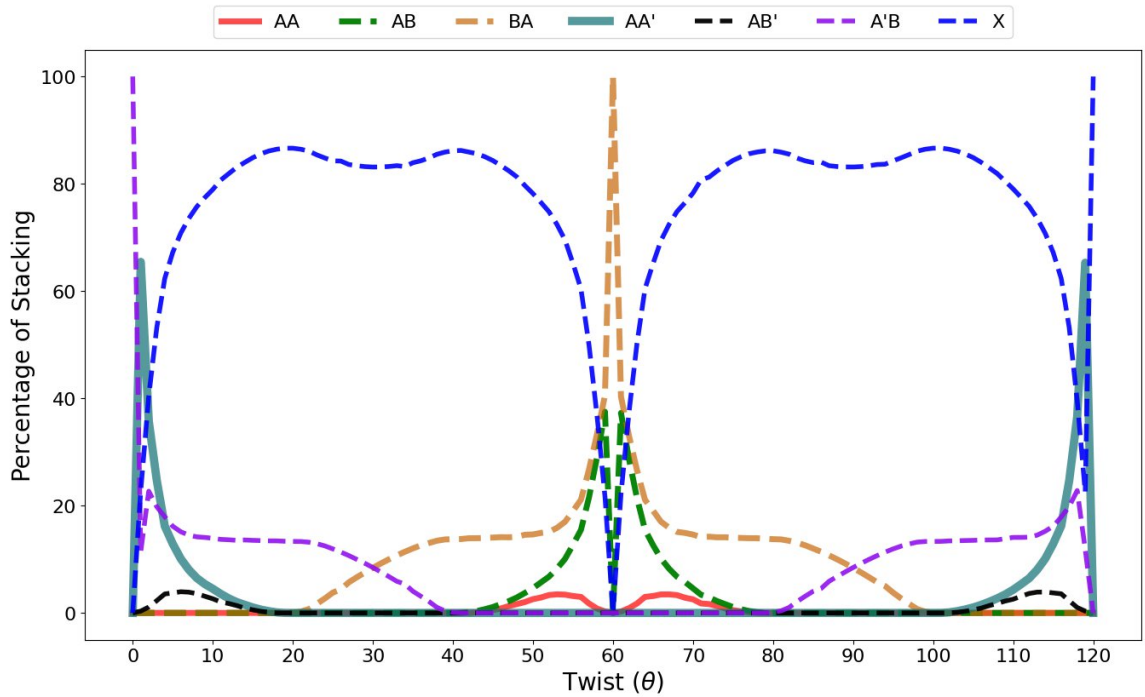

**Fig S6.** Starting with A'B stacked MoS$_2$ bilayers, twists are introduced between the bilayers and the percentage of atoms in different stacking types are calculated. At 60° rotation, the bilayers transform to BA and AA' stacking at 120°.